\documentclass[12pt]{article}

\usepackage{amsfonts}
\usepackage{amsmath}
\usepackage{color}
\usepackage{geometry}
\usepackage{graphicx} 
\begin{document}

\begin{center}
\textbf{Power new generalized class of Kavya--Manoharan distributions with
an application to exponential distribution}\bigskip

Lazhar Benkhelifa\bigskip

\textit{Department of Mathematics, Mohamed Khider University, Biskra, Algeria%
}\bigskip

lazhar.benkhelifa@univ-biskra.dz\bigskip \bigskip

\noindent\textbf{Abstract}\medskip
\end{center}

Recently, Verma et al. (2025) introduced a novel generalized class of
Kavya-Manoharan distributions, which have demonstrated significant utility
in reliability analysis and the modeling of lifetime data. This paper
proposes an extension of this class by applying the power generalization
technique, thereby enhancing more flexibility and applicability. We take the
exponential distribution as the baseline distribution to introduce a new
model capable of accommodating both monotonic and non-monotonic hazard rate
functions. Our model includes eleven submodels. We present several
statistical properties of the introduced model, including moments,
generating and characteristic functions, mean deviations, quantile function,
mean residual life function, R\'{e}nyi entropy, order statistics, and
reliability. To estimate the unknown model parameters, we use the maximum
likelihood approach. A simulation study is conducted to assess the validity
of the maximum likelihood estimator. The superiority of the new distribution
is demonstrated through the use of a real data application.\bigskip

\bigskip

\section{\textbf{Introduction}}

In reliability, lifetime models are important tools for modeling and
analyzing the time until a failure happens. The exponential distribution has
long been a popular choice for modeling lifetime data due to its simplicity
and ease of use. But it's a notable limitation when we want to model the
data that does not exhibit a constant hazard rate. In order to get rid of
this limitation, researchers have suggested several extensions and
generalizations of the exponential distribution via transformation methods.
The aim of these generalizations is to provide greater flexibility to model
various types of data by introducing at least one parameter or transforming
the parent distribution. For example, Gupta et al. (1998) used the power
generalization technique to introduce the exponentiated exponential
distribution. Nadarajah and Kotz (2006) suggested another generalization of
the exponential distribution, derived from the logit transformation of a
beta random variable, which they termed the beta exponential distribution.
Kumar et al. (2015) presented the Dinesh-Umesh-Sanjay (DUS) transformation
and studied the DUS exponential distribution. The alpha power transformation
(APT), with an application to exponential distribution, was developed by
Mahdavi and Kundu (2017). Kavya and Manoharan (2021) introduced the
Kavya-Manoharan (KM) transformation method, whereas the power generalized
DUS (PGDUS) transformation was proposed by Thomas and Chacko (2021). Lone
and Jan (2023) introduced the method known as the pi-exponentiated
transformation (PET). The power generalized KM (PGKM) transformation for
non-monotone hazard rate was presented by Deepthi and Chacko (2023). More
recently, by extending the KM transformation, Verma et al. (2025) developed
a new generalized KM (NGKM) transformation technique. They introduced the
NGKM exponential distribution utilizing the exponential distribution as the
baseline model. The cumulative distribution function (CDF) of the NGKM
transformation is given by: 
\begin{equation*}
F\left( x\right) =\left\{ 
\begin{tabular}{ll}
$\frac{\lambda }{\lambda -1}\left[ 1-\lambda ^{-G\left( x\right) }\right] $
& if $\lambda $ $>0,\lambda \neq 1,$\bigskip \\ 
$G\left( x\right) $ & if $\lambda $ $=1,$%
\end{tabular}%
\ \ \ \ \right.
\end{equation*}%
where $G(x)$ is the CDF of the baseline distribution and $\lambda $ is the
shape parameter.\medskip

In this paper, we apply the power generalization technique to propose a new
extension of the NGKM transformation, referred to as the power NGKM (PNGKM)
transformation. Adding a new shape parameter $\alpha >0,$ we get the CDF of
the PNGKM family, which is:%
\begin{equation}
F\left( x\right) =\left\{ 
\begin{tabular}{ll}
$\frac{\lambda ^{\alpha }}{\left( \lambda -1\right) ^{\alpha }}\left[
1-\lambda ^{-G\left( x\right) }\right] ^{\alpha }$ & if $\lambda $ $%
>0,\lambda \neq 1,$\bigskip \\ 
$G^{\alpha }\left( x\right) $ & if $\lambda $ $=1.$%
\end{tabular}%
\right.  \label{1.1}
\end{equation}%
\bigskip

Note that the PNGKM transformation contains some transformations techniques
as pecial cases:\medskip

\begin{itemize}
\item If $\lambda =1$, then we obtain the exponentiated (power)
transformation; which is proposed by Gupta et al (1998).

\item If \ $\alpha =1$ and $\lambda =e^{-1}$\textbf{,} then we get the DUS
transformation; see Kumar et al. (2015).

\item If $\alpha =1$ and $\lambda =\theta ^{-1}$, then we obtain the APT;
see Mahdavi and Kundu 2017.

\item If $\alpha =1$ and $\lambda =e$, then we get the GKM transformation;
see Kavya and Manoharan (2021).

\item If $\lambda =e$, then we obtain the PGKM transformation; see Deepthi
and Chacko (2023).

\item If \textbf{\ }$\lambda =e^{-1}$\textbf{, }then we get the PDUS; see
Thomas and Chacko (2023).

\item When $\alpha =1,$ $\lambda =\pi ^{-1}\ $and $\gamma =1$, we obtain the
PET; see Lone and Jan (2023).

\item When $\alpha =1$, we obtain the NGKM transformation; see Verma et al.
(2025).

\item When $\lambda =\theta ^{-1}$, we get the power APT (PAPT); new.

\item When $\lambda =\pi ^{-1}\ $and $\gamma =1$ we get the power PET
(PPET); new.\medskip \medskip
\end{itemize}

The paper is structured as follows. Section 2 presents the PNGKM exponential
distribution. Various statistical properties of our suggested distribution
are given in Section 3. The maximum likelihood estimators of the model
parameters and the examination of the performance of these estimators
through a simulation are presented in sections 4 and 5, respectively. An
application to well-known real data is given in Section 6. Finally, Section
7 was devoted to the conclusion..

\section{ \textbf{PNGKM exponential distribution}}

\noindent It is known that the CDF of exponential distribution is $G\left(
x\right) =1-e^{-\beta x},$ $\beta >0,\ x>0.$ Then, substituting $G\ $in $%
\left( 1\right) $ we get the CDF of the PNGKM exponential (PNGKME)\textbf{\ }%
distribution as follows:%
\begin{equation}
F\left( x\right) =\left\{ 
\begin{tabular}{ll}
$\frac{\lambda ^{\alpha }}{\left( \lambda -1\right) ^{\alpha }}\left(
1-\lambda ^{-\left( 1-e^{-\beta x}\right) }\right) ^{\alpha }$ & if $\lambda 
$ $>0,\lambda \neq 1,$\bigskip \\ 
$\left( 1-e^{-\beta x}\right) ^{\alpha }$ & if $\lambda $ $=1.$%
\end{tabular}%
\ \ \ \ \ \ \ \right.  \label{1}
\end{equation}%
The corresponding probability density function (PDF) and the hazard rate
function are, respectively, given by%
\begin{equation}
f\left( x\right) =\left\{ 
\begin{tabular}{ll}
$\frac{\alpha \beta \lambda ^{\alpha }\ln \lambda }{\left( \lambda -1\right)
^{\alpha }}e^{-\beta x}\lambda ^{-\left( 1-e^{-\beta x}\right) }\left(
1-\lambda ^{-\left( 1-e^{-\beta x}\right) }\right) ^{\alpha -1}$ & if $%
\lambda $ $>0,\lambda \neq 1,$\bigskip \\ 
$\alpha \beta e^{-\beta x}\left( 1-e^{-\beta x}\right) ^{\alpha -1}$ & if $%
\lambda $ $=1,$%
\end{tabular}%
\ \ \ \ \ \ \ \right.  \label{2}
\end{equation}%
and%
\begin{equation*}
h\left( x\right) =\left\{ 
\begin{tabular}{ll}
$\frac{\alpha \beta \lambda ^{\alpha }\ln \left( \lambda \right) e^{-\beta
x}\lambda ^{-\left( 1-e^{-\beta x}\right) }\left( 1-\lambda ^{-\left(
1-e^{-\beta x}\right) }\right) ^{\alpha -1}}{\left( \lambda -1\right)
^{\alpha }-\lambda ^{\alpha }\left( 1-\lambda ^{-\left( 1-e^{-\beta
x}\right) }\right) ^{\alpha }}$ & if $\lambda $ $>0,\lambda \neq 1,$\bigskip
\\ 
$\frac{\alpha \beta e^{-\beta x}\left( 1-e^{-\beta x}\right) ^{\alpha -1}}{%
1-\left( 1-e^{-\beta x}\right) ^{\alpha }}$ & if $\lambda $ $=1.$%
\end{tabular}%
\ \ \ \ \ \ \ \right.
\end{equation*}%
\bigskip

The PNGKME distribution includes as special cases many important well-known
distributions:\medskip

\begin{itemize}
\item When $\lambda =1$, it reduces to exponentiated exponential (EE)$\ $%
distribution; see Gupta et al (1998).

\item When $\alpha =1$ and $\lambda =e^{-1}$\textbf{,} we get the DUS
exponential (DUSE)$\ $distribution; see Kumar et al. (2015).

\item When $\alpha =1$ and $\lambda =\theta ^{-1}$, we obtain the APT
exponential (APTE)$;$ see Mahdavi and Kundu (2017).

\item When $\alpha =1$ and $\lambda =e$, it reduces to GKME distribution;
see Kavya and Manoharan (2021).

\item When $\lambda =e$, it reduces to PGKME distribution; see Deepthi and
Chacko (2023).

\item If \textbf{\ }$\lambda =e^{-1}$\textbf{, }then\textbf{\ }we obtain the
PDUSE distribution; see Thomas and Chacko (2021).

\item When $\alpha =1,$ $\lambda =\pi ^{-1}\ $and $\gamma =1$, we get the
PETE distribution; see Lone and Jan (2023).

\item When $\alpha =1$, we get the NGKME distribution; see Verma et al.
(2025).

\item When $\lambda =\theta ^{-1}$, we obtain the PAPT exponential (PAPTE)$\ 
$distribution; new.

\item If $\lambda =\pi ^{-1}$ and $\gamma =1$, then it reduces to PPET
exponential (PPETE)$\ $distribution; new.

\item If $\lambda =1$ and $\alpha =1,$ then we get the exponential$\ $%
distribution.\bigskip
\end{itemize}

The general shapes of the PDF and hazard rate function for the PNGKME
distribution are difficult, or even impossible, to establish analytically.
Nonetheless, we can graphically illustrate these shapes for some selected
values of $\alpha ,$ $\beta $ and $\lambda $. Figure 1 gives the various
shapes of the PDF of the PNGKME distribution. From this figure, it is
evident that the PDF has a decreasing or a unimodal shape depending on the
parameter values.\bigskip

Similarly, various possible shapes of the hazard function of the PNGKME
distribution are given in Figure 2. We observe that the shapes of hazard
rates are decreasing, increasing, upside-down bathtub (unimodal), or
constant according to the choice of parameters.

\begin{figure}[!htb]
\centering
\includegraphics[height=5cm,width=10cm]{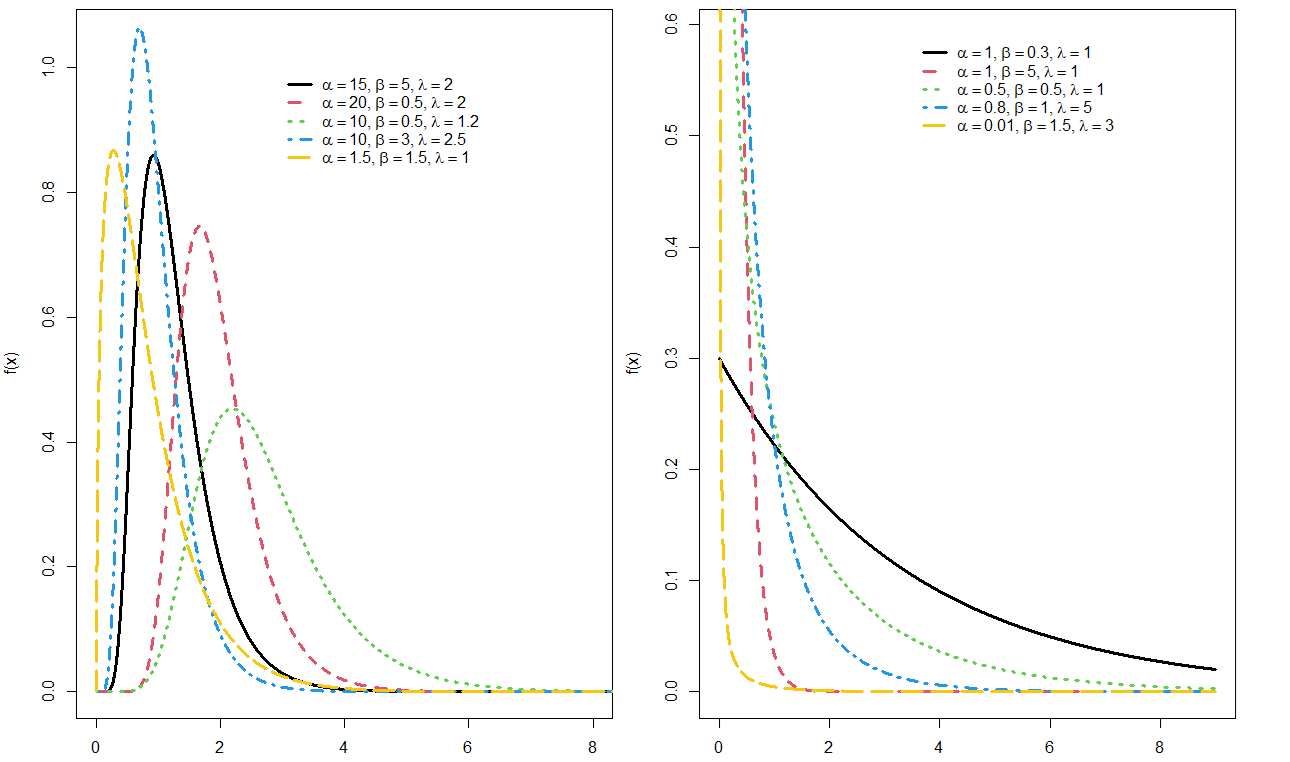}
\caption{PDF plot of the PNGKME distribution.}
\end{figure}

\begin{figure}[!htb]
\centering
\includegraphics[height=5cm,width=10cm]{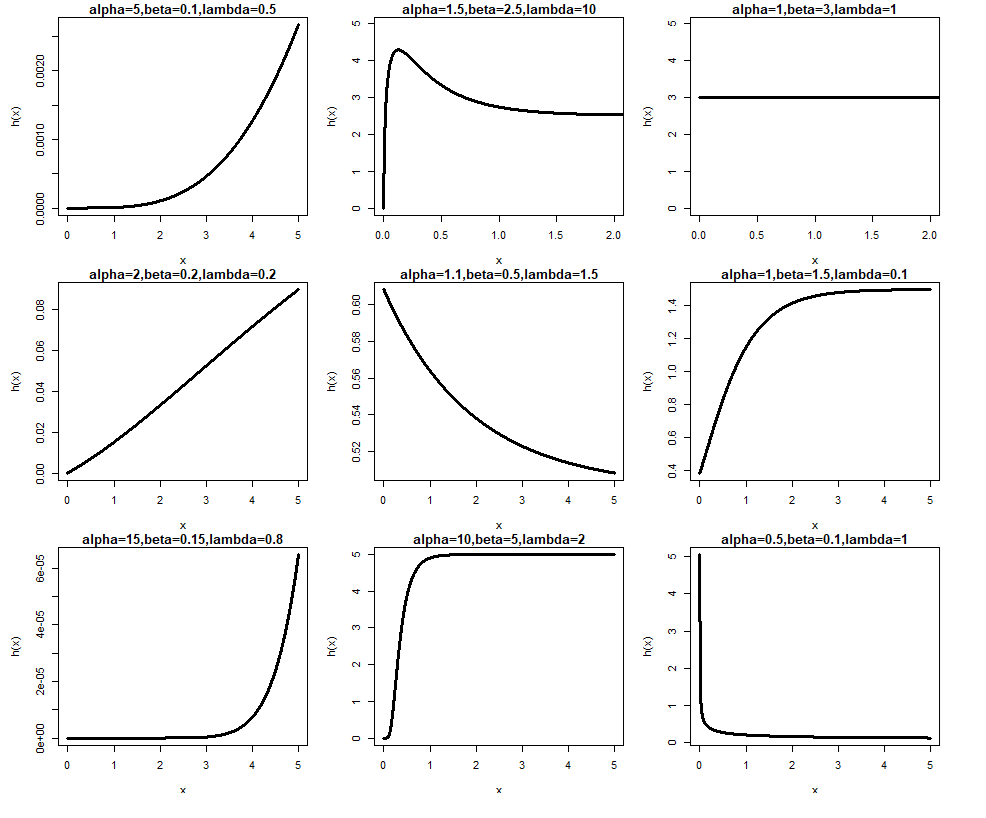}
\caption{Some possible shapes of the hazard function}
\end{figure}

\section{\textbf{Statistical properties}}

Note that a survey of the properties of the EE distribution for $\lambda $ $%
=1$ was studied by Nadarajah (2011), while in this section we present
nemerous statistical properties of our introduced PNGKME distribution, for $%
\lambda $ $>0,\ $where $\lambda \neq 1$.\bigskip

\subsection{Moments}

Let $X$ denote a random variable having the PDF (3). Then, the $r$th moment
about the origin of $X$ is%
\begin{equation*}
E\left( X^{r}\right) =\frac{\alpha \beta \lambda ^{\alpha }\ln \lambda }{%
\left( \lambda -1\right) ^{\alpha }}\int_{0}^{+\infty }x^{r}e^{-\beta
x}\lambda ^{-\left( 1-e^{-\beta x}\right) }\left( 1-\lambda ^{-\left(
1-e^{-\beta x}\right) }\right) ^{\alpha -1}dx.
\end{equation*}%
Making use of the following series%
\begin{equation}
\left( 1-z\right) ^{\delta }=\sum_{j=0}^{\infty }\left( -1\right) ^{j}\binom{%
\delta }{j}z^{j},\text{ for }\left\vert z\right\vert <1,\text{ }\delta >0,
\label{3}
\end{equation}%
we get for $\lambda $ $>0\ $and $\lambda \neq 1$%
\begin{equation*}
E\left( X^{r}\right) =\frac{\alpha \beta \lambda ^{\alpha }\ln \lambda }{%
\left( \lambda -1\right) ^{\alpha }}\sum_{j=0}^{\infty }\left( -1\right) ^{j}%
\binom{\alpha -1}{j}\lambda ^{-\left( j+1\right) }\int_{0}^{+\infty
}x^{r}e^{-\beta x}\lambda ^{\left( j+1\right) e^{-\beta x}}dx,
\end{equation*}%
and then using the following series representation%
\begin{equation}
\lambda ^{u}=\sum_{k=0}^{\infty }\frac{u^{k}\left( \ln \lambda \right) ^{k}}{%
k!},  \label{4}
\end{equation}%
we obtain 
\begin{equation*}
E\left( X^{r}\right) =\frac{\alpha r!}{\beta ^{r}\left( \lambda -1\right)
^{\alpha }}\sum_{j=0}^{\infty }\sum_{k=1}^{\infty }\left( -1\right) ^{j}%
\binom{\alpha -1}{j}\frac{\left( j+1\right) ^{k-1}\left( \ln \lambda \right)
^{k}}{k!k^{r}\lambda ^{\left( j+1\right) -\alpha }}.
\end{equation*}

\subsection{Generating, characteristic and cumulant generating functions}

The generating function of the PNGKME$\ $distribution is%
\begin{eqnarray*}
M_{X}(t) &=&E\left( e^{tX}\right) \\
&=&\frac{\alpha \beta \lambda ^{\alpha }\ln \lambda }{\left( \lambda
-1\right) ^{\alpha }}\int_{0}^{+\infty }e^{tx}e^{-\beta x}\lambda ^{-\left(
1-e^{-\beta x}\right) }\left( 1-\lambda ^{-\left( 1-e^{-\beta x}\right)
}\right) ^{\alpha -1}dx.
\end{eqnarray*}%
Using (4) and (5), and after some algebra, we obtain%
\begin{equation*}
M_{X}(t)=\frac{\alpha \beta }{\left( \lambda -1\right) ^{\alpha }}%
\sum_{j=0}^{\infty }\sum_{k=1}^{\infty }\left( -1\right) ^{j}\binom{\alpha -1%
}{j}\frac{\lambda ^{\alpha -\left( j+1\right) }\left( j+1\right)
^{k-1}\left( \ln \lambda \right) ^{k}}{\left( k-1\right) !\left( k\beta
-t\right) },\text{ \ }\beta <t.
\end{equation*}%
Similarly, one can get the characteristic function as follows%
\begin{eqnarray*}
\phi (t) &=&E\left( e^{itX}\right) \\
&=&\frac{\alpha \beta }{\left( \lambda -1\right) ^{\alpha }}%
\sum_{j=0}^{\infty }\sum_{k=1}^{\infty }\left( -1\right) ^{j}\binom{\alpha -1%
}{j}\frac{\lambda ^{\alpha -\left( j+1\right) }\left( j+1\right)
^{k-1}\left( \ln \lambda \right) ^{k}}{\left( k-1\right) !\left( k\beta
-it\right) },
\end{eqnarray*}%
whereas the cumulant generating function is 
\begin{eqnarray*}
K(t) &=&\ln \phi (t) \\
&=&\ln \left( \frac{\alpha \beta }{\left( \lambda -1\right) ^{\alpha }}%
\right) +\ln \left( \sum_{j=0}^{\infty }\sum_{k=1}^{\infty }\left( -1\right)
^{j}\binom{\alpha -1}{j}\frac{\lambda ^{\alpha -\left( j+1\right) }\left(
j+1\right) ^{k-1}\left( \ln \lambda \right) ^{k}}{\left( k-1\right) !\left(
k\beta -it\right) }\right) ,
\end{eqnarray*}%
where $i=\sqrt{-1}$.

\subsection{\textbf{Quantile function}}

To find the quantile function $Q(u)$, we must find the root of the equation $%
F(Q(u))=u,$ where $F$ is the CDF given in (2). After some simple
computation, we get%
\begin{equation}
Q(u)=\frac{-1}{\beta }\ln \left( 1+\frac{\ln \left( 1-\left( \frac{\left(
\lambda -1\right) ^{\alpha }}{\lambda ^{\alpha }}u\right) ^{1/\alpha
}\right) }{\ln \lambda }\right) ,\text{ for }u\in \left[ 0,1\right] .
\label{5}
\end{equation}%
Therefore the random variable $X$ given by%
\begin{equation}
X=\ \frac{-1}{\beta }\ln \left( 1+\frac{\ln \left( 1-\left( \frac{\left(
\lambda -1\right) ^{\alpha }}{\lambda ^{\alpha }}U\right) ^{1/\alpha
}\right) }{\ln \lambda }\right)  \label{6}
\end{equation}%
follows the PNGKME distribution, where $U$ is a continuous uniform variable
on $\left[ 0,1\right] .$\bigskip

\noindent Setting $u=0.5$ in (6), we get the median of the PNGKME
distribution, which is given by 
\begin{equation*}
M=Q\left( \frac{1}{2}\right) =\ \frac{-1}{\beta }\ln \left( 1+\frac{\ln
\left( 1-\left( \frac{\left( \lambda -1\right) ^{\alpha }}{2\lambda ^{\alpha
}}\right) ^{1/\alpha }\right) }{\ln \lambda }\right) ,
\end{equation*}%
while by setting $u=0.25$ and $u=0.75$ in (6), we get the first quartile and
third quartile, respectively.

\bigskip

\subsection{\textbf{Mean residual life function}}

In reliability, the mean residual life function is an important
characteristic of the model. It is defined as the expected additional
lifetime $(X-t)$ for $t>0,$ of a component or system given that the
component or the system has survived till time $t$. The mean residual life
function of the PNGKME distribution is%
\begin{eqnarray*}
E\left( X-t\right\vert X &>&t)=\frac{\int_{0}^{\infty }xf\left( x\right) dx}{%
1-F\left( t\right) }-t \\
&=&\frac{\alpha \sum_{j=0}^{\infty }\sum_{k=1}^{\infty }\left( -1\right) ^{j}%
\binom{\alpha -1}{j}\frac{\left( j+1\right) ^{k-1}\left( \ln \lambda \right)
^{k}}{k!k\lambda ^{\left( j+1\right) -\alpha }}}{\beta \left( \lambda
-1\right) ^{\alpha }-\beta ^{r}\lambda ^{\alpha }\left( 1-\lambda ^{-\left(
1-e^{-\beta t}\right) }\right) ^{\alpha }}-t.
\end{eqnarray*}

\subsection{\textbf{Mean deviations}}

\noindent The mean absolute deviation of any random variable $X$ from its
mean $\mu =E(X)$ is%
\begin{equation*}
\delta _{1}=E\left( \left\vert X-\mu \right\vert \right) =2\mu F\left( \mu
\right) -2\mu +2\int_{\mu }^{\infty }xf\left( x\right) dx,
\end{equation*}%
while the mean absolute deviation of $X$ from its median $M=Q\left(
1/2\right) $, is%
\begin{equation*}
\delta _{2}=E\left( \left\vert X-M\right\vert \right) =-\mu
+2\int_{M}^{\infty }xf\left( x\right) dx,
\end{equation*}%
If $X$ has the PDF (3), therefore after some algebra, we obtain%
\begin{eqnarray*}
\delta _{1} &=&2\mu F\left( \mu \right) -2\mu +\frac{2\alpha }{\beta \left(
\lambda -1\right) ^{\alpha }} \\
&&\times \sum_{j=0}^{\infty }\sum_{k=1}^{\infty }\left( -1\right) ^{j}\binom{%
\alpha -1}{j}\frac{\left( j+1\right) ^{k-1}\left( \ln \lambda \right)
^{k}\lambda ^{\alpha -\left( j+1\right) }}{k!k}\left( k\beta \mu +1\right)
e^{-k\beta \mu },
\end{eqnarray*}%
and%
\begin{eqnarray*}
\delta _{2} &=&-\mu +\frac{2\alpha }{\beta \left( \lambda -1\right) ^{\alpha
}} \\
&&\times \sum_{j=0}^{\infty }\sum_{k=1}^{\infty }\left( -1\right) ^{j}\binom{%
\alpha -1}{j}\frac{\left( j+1\right) ^{k-1}\left( \ln \lambda \right)
^{k}\lambda ^{\alpha -\left( j+1\right) }}{k!k}\left( k\beta M+1\right)
e^{-k\beta M}.
\end{eqnarray*}

\subsection{\textbf{Bonferroni and Lorenz curves}}

\noindent These curves are used in several domains like reliability,
medicine and economics. The Bonferroni and Lorenz curves are given by%
\begin{equation*}
B\left( p\right) =\frac{1}{p\mu }\int_{0}^{q}xf\left( x\right) dx\text{ and }%
L\left( p\right) =\frac{1}{\mu }\int_{0}^{q}xf\left( x\right) dx,
\end{equation*}%
respectively, where $\mu =E(X)$ and $q=Q(p)$. We can rewrite these curves as
follows%
\begin{equation*}
B\left( p\right) =\frac{1}{p}-\frac{1}{p\mu }\int_{q}^{\infty }xf\left(
x\right) dx\text{ and }L\left( p\right) =1-\frac{1}{\mu }\int_{q}^{\infty
}xf\left( x\right) dx.
\end{equation*}%
Therefore, if $X$ has the PDF (3), we get 
\begin{equation*}
B\left( p\right) =\frac{1}{p}-\sum_{j=0}^{\infty }\sum_{k=1}^{\infty }\left(
-1\right) ^{j}\binom{\alpha -1}{j}\frac{\left( j+1\right) ^{k-1}\left( \ln
\lambda \right) ^{k}\lambda ^{\alpha -\left( j+1\right) }\left( k\beta
q+1\right) \alpha e^{-k\beta q}}{p\mu \beta \left( \lambda -1\right)
^{\alpha }k!k},
\end{equation*}%
and%
\begin{equation*}
L\left( p\right) =1-\sum_{j=0}^{\infty }\sum_{k=1}^{\infty }\left( -1\right)
^{j}\binom{\alpha -1}{j}\frac{\left( j+1\right) ^{k-1}\left( \ln \lambda
\right) ^{k}\lambda ^{\alpha -\left( j+1\right) }\left( k\beta \mu +1\right)
\alpha e^{-k\beta \mu }}{\mu \beta \left( \lambda -1\right) ^{\alpha }k!k}.
\end{equation*}

\subsection{\textbf{R\'{e}nyi entropy}}

\noindent Entropy, which has applications in a wide variety domains,
measures the degree of uncertainty associated with a random variable.
According to R\'{e}nyi (1961), the R\'{e}nyi entropy is%
\begin{equation*}
I_{R}\left( s\right) =\frac{1}{1-s}\log \left( \int_{%
\mathbb{R}
}f^{s}\left( x\right) dx\right) ,\text{ }s>0,\text{ }s\neq 1.
\end{equation*}%
When $s\rightarrow 1,$ we obtain the Shannon entropy (Shannon, 1951). Then,
if $X$ follows the PDF (2), we have%
\begin{equation*}
I_{R}\left( s\right) =\frac{1}{1-s}\log \left( \int_{0}^{\infty }f^{s}\left(
x\right) dx\right) ,\text{ }s>0,\text{ }s\neq 1,
\end{equation*}%
where%
\begin{equation*}
f^{s}\left( x\right) =%
\begin{tabular}{ll}
$\left( \frac{\alpha \beta \lambda ^{\alpha }\ln \lambda }{\left( \lambda
-1\right) ^{\alpha }}\right) ^{s}e^{-s\beta x}\lambda ^{-s\left( 1-e^{-\beta
x}\right) }\left( 1-\lambda ^{-\left( 1-e^{-\beta x}\right) }\right)
^{s\left( \alpha -1\right) }$ & if $\lambda $ $>0,\lambda \neq 1,$%
\end{tabular}%
.
\end{equation*}%
by applying (4) and (5), and after some algebra, we get%
\begin{eqnarray*}
I_{R}\left( s\right) &=&\frac{s}{1-s}\ln \left( \frac{\alpha \beta \lambda
^{\alpha }\ln \lambda }{\left( \lambda -1\right) ^{\alpha }}\right) \\
&&+\frac{s}{1-s}\ln \left( \sum_{j=0}^{\infty }\sum_{k=0}^{\infty }\left(
-1\right) ^{j}\binom{s\left( \alpha -1\right) }{j}\frac{\lambda ^{-\left(
s+j\right) }\left( s+j\right) ^{k}\left( \ln \lambda \right) ^{k}}{k!\beta
\left( s+k\right) }\right) .
\end{eqnarray*}%
\qquad \qquad \qquad \qquad \qquad \qquad \qquad \qquad \qquad \qquad \qquad
\qquad \qquad \qquad \qquad \qquad \qquad \qquad \qquad

\subsection{\textbf{Reliability}}

\noindent The stress-strength model $R=\mathbb{P}(X_{2}<X_{1})$ is used to
evaluate the reliability of a system or component when it is subjected to
stress $X_{2}$ and has strength $X_{1}.$ The system functions properly when
the strength exceeds the stress, $X_{2}<X_{1}.$ To compute $R$, we
distinguish four cases in which $X_{1}$ and $X_{2}$ have distinct PNGKME
distributions, where 
\begin{equation*}
R=\int_{0}^{\infty }f\left( x;\alpha ,\beta ,\lambda _{1}\right) F\left(
;\alpha ,\beta ,\lambda _{2}\right) dx.
\end{equation*}

\noindent \underline{\textit{Case 1: }$\lambda _{1}$\textit{\ }$\neq 1$%
\textit{\ and }$\lambda _{2}$\textit{\ }$\neq 1$}%
\begin{equation*}
R=\int_{0}^{\infty }\frac{\alpha \beta \lambda _{1}^{\alpha }\ln \lambda _{1}%
}{\left( \lambda _{1}-1\right) ^{\alpha }}\frac{\lambda _{2}^{\alpha }}{%
\left( \lambda _{2}-1\right) ^{\alpha }}e^{-\beta x}\lambda _{1}^{-\left(
1-e^{-\beta x}\right) }\left( 1-\lambda _{1}^{-\left( 1-e^{-\beta x}\right)
}\right) ^{\alpha -1}\left( 1-\lambda _{2}^{-\left( 1-e^{-\beta x}\right)
}\right) ^{\alpha }dx
\end{equation*}%
Using (4) and (5), and after some algebra, we get%
\begin{equation*}
R=\frac{\alpha \lambda _{1}^{\alpha }\ln \lambda _{1}}{\left( \lambda
_{1}-1\right) ^{\alpha }}\frac{\lambda _{2}^{\alpha }}{\left( \lambda
_{2}-1\right) ^{\alpha }}\sum_{j=0}^{\infty }\sum_{i=0}^{\infty }\left(
-1\right) ^{i+j}\binom{\alpha }{i}\binom{\alpha -1}{j}\frac{1-\lambda
_{1}^{-\left( i+j\right) -\frac{i\ln \lambda _{2}}{\ln \lambda _{1}}}}{i\ln
\lambda _{2}+\left( i+j\right) \ln \lambda _{1}}
\end{equation*}

\noindent \underline{\textit{Case 2: }$\lambda _{1}$\textit{\ }$=1$\textit{\
and }$\lambda _{2}$\textit{\ }$\neq 1$}%
\begin{equation*}
R=\int_{0}^{\infty }\alpha \beta e^{-\beta x}\left( 1-e^{-\beta x}\right)
^{\alpha -1}\frac{\lambda _{2}^{\alpha }}{\left( \lambda _{2}-1\right)
^{\alpha }}\left( 1-\lambda _{2}^{-\left( 1-e^{-\beta x}\right) }\right) dx,
\end{equation*}%
by taking $z=\left( 1-e^{-\beta x}\right) ,$ we get%
\begin{eqnarray*}
R &=&\frac{\lambda _{2}^{\alpha }\alpha }{\left( \lambda _{2}-1\right)
^{\alpha }}\int_{0}^{1}z^{\alpha -1}\left( 1-\lambda _{2}^{-m}\right) dz \\
&=&\frac{\lambda _{2}^{\alpha }\left( \left( \ln \lambda _{2}\right)
^{\alpha }+\alpha \Gamma \left( \alpha ,\ln \lambda _{2}\right) -\alpha
\Gamma \left( \alpha \right) \right) }{\left( \lambda _{2}-1\right) ^{\alpha
}\left( \ln \lambda _{2}\right) ^{\alpha }},
\end{eqnarray*}%
where $\Gamma $ denotes gamma function.\bigskip

\noindent \underline{\textit{Case 3: }$\lambda _{1}$\textit{\ }$\neq 1$%
\textit{\ and }$\lambda _{2}$\textit{\ }$=1$}%
\begin{equation*}
R=\int_{0}^{\infty }\frac{\alpha \beta \lambda _{1}^{\alpha }\ln \lambda _{1}%
}{\left( \lambda _{1}-1\right) ^{\alpha }}e^{-\beta x}\lambda _{1}^{-\left(
1-e^{-\beta x}\right) }\left( 1-\lambda _{1}^{-\left( 1-e^{-\beta x}\right)
}\right) ^{\alpha -1}\left( 1-e^{-\beta x}\right) ^{\alpha }dx.
\end{equation*}%
Using (4) and taking $z=\left( 1-e^{-\beta x}\right) ,$ we get%
\begin{equation*}
R=\frac{\alpha \lambda _{1}^{\alpha }\ln \lambda _{1}}{\left( \lambda
_{1}-1\right) ^{\alpha }}\sum_{j=0}^{\infty }\left( -1\right) ^{j}\binom{%
\alpha -1}{j}\frac{\Gamma \left( \alpha +1,0\right) -\Gamma \left( \alpha
+1,\left( j+1\right) \ln \lambda _{1}\right) }{\left( j+1\right) \ln \lambda
_{1}\left[ \left( j+1\right) \ln \lambda _{1}\right] ^{\alpha }},
\end{equation*}%
where $\Gamma $ denotes gamma function.\bigskip

\noindent \underline{\textit{Case 4: }$\lambda _{1}$\textit{\ }$=1$\textit{\
and }$\lambda _{2}$\textit{\ }$=1$}%
\begin{equation*}
R=\int_{0}^{\infty }\alpha _{1}\beta _{1}e^{-\beta _{1}x}\left( 1-e^{-\beta
_{1}x}\right) ^{\alpha _{1}-1}\left( 1-e^{-\beta _{2}x}\right) ^{\alpha
_{2}}dx.
\end{equation*}%
Using (4), and after some computations, we get%
\begin{equation*}
R=\sum_{j=0}^{\infty }\sum_{k=0}^{\infty }\binom{\alpha _{1}-1}{j}\binom{%
\alpha _{2}}{k}\frac{\left( -1\right) ^{j+k}\alpha _{1}\beta _{1}}{\left(
j+1\right) \beta _{1}+k\beta _{2}}.
\end{equation*}

\subsection{\textbf{Order statistics}}

In reliability and life testing, the order statistics play a crucial role.
They provide a mathematical framework for analyzing the lifetimes of systems
or components, especially when dealing with extremes (minimum or maximum
lifetimes) or rankings in a sample. From Arnold et al. (2008), the $k^{th}$
order statistic $X_{k,n}$ has the following CDF%
\begin{equation*}
F_{k,n}\left( x\right) =\sum_{j=k}^{n}\sum_{l=0}^{n-j}\left( -1\right) ^{l}%
\binom{n}{j}\binom{n-j}{l}F\left( x\right) ^{j+l},\text{ for }k=1,\ldots ,n,
\end{equation*}
\ and the following PDF%
\begin{equation*}
f_{k,n}\left( x\right) =\frac{n!}{\left( n-k\right) !\left( k-1\right) !}%
\sum_{l=0}^{n-k}\left( -1\right) ^{l}\binom{n-k}{l}f\left( x\right) \left[
F\left( x\right) \right] ^{k+l-1}.
\end{equation*}
So, the CDF of $X_{k,n}$ from the sample $X_{1},\ldots ,X_{n}$ having the
PNGKME distribution is given by 
\begin{equation*}
F_{k,n}\left( x\right) =\sum_{j=k}^{n}\sum_{l=0}^{n-j}\left( -1\right) ^{l}%
\binom{n}{j}\binom{n-j}{l}\frac{\lambda ^{\alpha \left( j+l\right) }}{\left(
\lambda -1\right) ^{\alpha \left( j+l\right) }}\left( 1-\lambda ^{-\left(
1-e^{-\beta x}\right) }\right) ^{\alpha \left( j+l\right) },
\end{equation*}%
whereas its PDF is%
\begin{equation*}
f_{k,n}\left( x\right) =\frac{n!\alpha \beta \lambda ^{\alpha }\ln \lambda
e^{-\beta x}}{\left( n-k\right) !\left( k-1\right) !}\sum_{l=0}^{n-k}\left(
-1\right) ^{l}\binom{n-k}{l}\frac{\lambda ^{\alpha \left( k+l-1\right)
-\left( 1-e^{-\beta x}\right) }}{\left( \lambda -1\right) ^{\alpha \left(
k+l\right) }}\left( 1-\lambda ^{-\left( 1-e^{-\beta x}\right) }\right)
^{\alpha \left( k+l\right) -1},
\end{equation*}%
for $\lambda $ $>0,$ $\lambda \neq 1$\ and $k=1,\ldots ,n.$

\section{\textbf{Maximum Likelihood Estimates}}

\noindent Here, we give the maximum likelihood estimate (MLE) and the
approximate confidence intervals of the PNGKME model parameters $\alpha
,\beta \ $and $\lambda $. If $x_{1},\ldots ,x_{n}$ is an observed values of
a random sample of size $n$ drawn from PNGKME distribution, then the
likelihood function is%

\begin{equation}
L=\left( \frac{\alpha \beta \lambda ^{\alpha }\ln \lambda }{\left( \lambda-1\right) ^{\alpha }}\right) ^{n}\sum_{i=1}^{n}\left( e^{-\beta x_{i}}\lambda ^{-\left( 1-e^{-\beta x_{i}}\right) }\left( 1-\lambda ^{-\left( 1-e^{-\beta x_{i}}\right) }\right) ^{\alpha -1}\right).  \label{7}
\end{equation}

The log-likelihood function is obtained by taking the logarithm of $L$ in
(8):%
\begin{align*}
\ln L& =n\ln \alpha +n\ln \beta +n\alpha \ln \lambda +n\ln \ln \lambda
-\alpha n\ln \left( \lambda -1\right) -\beta \sum_{i=1}^{n}x_{i} \\
& -\left( n-\sum_{i=1}^{n}e^{-\beta x_{i}}\right) \ln \lambda +\left( \alpha
-1\right) \sum_{i=1}^{n}\ln \left( 1-\lambda ^{-\left( 1-e^{-\beta
x_{i}}\right) }\right) .
\end{align*}%
The partial derivatives of $\ln L$ w.r.t $\alpha ,\beta \ $and $\lambda $ are%
\begin{equation*}
\frac{\partial \ln L}{\partial \alpha }=\frac{n}{\alpha }+n\ln \lambda -n\ln
\left( \lambda -1\right) +\sum_{i=1}^{n}\ln \left( 1-\lambda ^{-\left(
1-e^{-\beta x_{i}}\right) }\right) ,
\end{equation*}%
\begin{equation*}
\frac{\partial \ln L}{\partial \beta }=\frac{n}{\beta }-\sum_{i=1}^{n}x_{i}-%
\ln \lambda \sum_{i=1}^{n}x_{i}e^{-\beta x_{i}}-\left( \alpha -1\right) \ln
\lambda \sum_{i=1}^{n}\frac{x_{i}\lambda ^{e^{-\beta x_{i}}}e^{-\beta x_{i}}%
}{\lambda ^{e^{-\beta x_{i}}}-\lambda },
\end{equation*}%
and%
\begin{equation*}
\frac{\partial \ln L}{\partial \lambda }=-\frac{n}{\lambda \ln \lambda }-%
\frac{\alpha n}{\lambda -1}+\frac{n\alpha -n-\sum_{i=1}^{n}e^{-\beta x_{i}}}{%
\lambda }+\left( \alpha -1\right) \sum_{i=1}^{n}\frac{\left( e^{-\beta
x_{i}}-1\right) \lambda ^{-\left( 2-e^{-\beta x_{i}}\right) }}{1-\lambda
^{-\left( 1-e^{-\beta x_{i}}\right) }}.
\end{equation*}

\noindent Solving the non-linear equations simultaneously $\frac{\partial
\ln L}{\partial \alpha }=0,$ $\frac{\partial \ln L}{\partial \beta }=0$ and $%
\frac{\partial \ln L}{\partial \lambda }=0,$ we get the MLE $\widehat{\alpha 
},\widehat{\beta }$ and $\widehat{\lambda }$ of $\alpha ,\beta \ $and $%
\lambda $ respectively. We use statistical software, like \textit{R}, to get
the solution of these equations numerically using iterative methods because
they cannot be solved analytically.\bigskip

To obtain the confidence intervals of $\alpha ,\beta \ $and $\lambda $, we
use the asymptotic distribution of their MLEs. This asymptotic distribution
is not normal, and the lower confidence limit could be negative, although $%
\alpha ,\beta \ $and $\lambda $ are positive (see Singh et al. 2013 and
Sharma et al. 2016). According to Burnham (1987), the log-based approximate $%
100(1-\omega )\%$ confidence intervals for $\alpha ,\beta \ $and $\lambda $
are%
\begin{equation*}
\alpha \in \left( \widehat{\alpha }/A,\widehat{\alpha }A\right) ,\text{ }%
\beta \in \left( \widehat{\beta }/B,\widehat{\beta }B\right) \ \text{and }%
\lambda \in \left( \widehat{\lambda }/C,\widehat{\lambda }C\right) ,
\end{equation*}%
where%
\begin{equation*}
A=\exp \left( Z_{\frac{\omega }{2}}\sqrt{\ln \left( 1+\frac{Var\left( 
\widehat{\alpha }\right) }{\widehat{\alpha }^{2}}\right) }\right) ,B=\exp
\left( Z_{\frac{\omega }{2}}\sqrt{\ln \left( 1+\frac{Var\left( \widehat{%
\beta }\right) }{\widehat{\beta }^{2}}\right) }\right) ,
\end{equation*}%
and%
\begin{equation*}
C=\exp \left( Z_{\frac{\omega }{2}}\sqrt{\ln \left( 1+\frac{Var\left( 
\widehat{\lambda }\right) }{\widehat{\lambda }^{2}}\right) }\right) ,
\end{equation*}%
whith $Z_{\frac{\omega }{2}}$ is the upper $100\times (\omega /2)^{th}$
percentile of a standard normal distribution and $Var\left( \widehat{\alpha }%
\right) ,$ $Var\left( \widehat{\beta }\right) $ and $Var\left( \widehat{%
\lambda }\right) $ are the variances of $\widehat{\alpha },\widehat{\beta }$
and $\widehat{\lambda }$ respectively. The variances are diagonal elements
of estimated variance-covariance, which is the inverse of Fisher's
information matrix:%
\begin{equation*}
-\left( 
\begin{array}{ccc}
\frac{\partial ^{2}\ln L}{\partial \alpha ^{2}} & \frac{\partial \ln L}{%
\partial \alpha \partial \beta } & \frac{\partial \ln L}{\partial \alpha
\partial \lambda } \\ 
\frac{\partial \ln L}{\partial \alpha \partial \beta } & \frac{\partial
^{2}\ln L}{\partial \beta ^{2}} & \frac{\partial \ln L}{\partial \beta
\partial \lambda } \\ 
\frac{\partial \ln L}{\partial \alpha \partial \lambda } & \frac{\partial
\ln L}{\partial \beta \partial \lambda } & \frac{\partial ^{2}\ln L}{%
\partial \lambda ^{2}}%
\end{array}%
\right) ,
\end{equation*}%
whose elements are given in the appendix.\bigskip

\section{Simulation analysis}

\noindent To examine the performance of the MLEs of $\alpha ,$ $\beta $ and $%
\lambda $, we conduct a simulation study. Using the equation (6), we
generate random samples from the PNGKME distribution. We repeat the
simulation $N=1000$ times for various sample sizes $n=50,$ $200,$ $500,$ $%
1000$. Four different sets of true values of $\alpha ,$ $\beta $ and $%
\lambda $ are considered. The performance of the MLEs is evaluated using
bias and mean squared error (MSE), which are given by:%
\begin{equation*}
Bias=\frac{1}{N}\sum_{i=1}^{N}\left( \widehat{\epsilon }_{i}-\epsilon
\right) \text{ \ and \ }MSE=\frac{1}{N}\sum_{i=1}^{N}\left( \widehat{%
\epsilon }_{i}-\epsilon \right) ^{2},
\end{equation*}

where $\epsilon =\left\{ \alpha ,\beta ,\lambda \right\} $ and $\widehat{%
\epsilon }_{i}$ is the MLE of $\epsilon $ in the $i$-th replication.\bigskip

Table 1 shows the results of our simulation study. From this table, it is
clear that the bias and MSE of the MLEs converges to $0$ as the sample size $%
n$ increases.. This confirms that the MLEs are consistent.

\begin{table}[h]%
\caption{Bias and MSE of $\protect\hat{\alpha}$, $\protect\hat{\beta}$ and
$\protect\hat{\lambda}$. }%
\begin{tabular}{c|ccc|ccc|ccc}
\hline
Sample size & \multicolumn{3}{|c}{Parameters} & \multicolumn{3}{|c}{Bias} & 
\multicolumn{3}{|c}{MSE} \\ \hline
$n$ & $\alpha $ & $\beta $ & $\lambda $ & $\widehat{\alpha }$ & $\widehat{%
\beta }$ & $\widehat{\lambda }$ & $\widehat{\alpha }$ & $\widehat{\beta }$ & 
$\widehat{\lambda }$ \\ \hline\hline
$50$ & $2.5$ & $1.5$ & $0.5$ & $0.0215$ & $0.2825$ & $0.2671$ & $0.4172$ & $%
0.7095$ & $0.7716$ \\ 
& $5$ & $2.5$ & $0.5$ & $0.1932$ & $0.1677$ & $0.5891$ & $1.5215$ & $1.7135$
& $0.9416$ \\ 
& $3.5$ & $5.5$ & $1.5$ & $0.1560$ & $0.1172$ & $0.7994$ & $0.8759$ & $%
1.3010 $ & $1.5611$ \\ 
& $1$ & $1.5$ & $2$ & $0.1030$ & $0.0617$ & $0.7131$ & $0.4382$ & $1.6291$ & 
$0.7865$ \\ \hline
$200$ & $2.5$ & $1.5$ & $0.5$ & $0.0023$ & $0.1334$ & $0.1230$ & $0.2753$ & $%
0.3465$ & $0.5562$ \\ 
& $5$ & $2.5$ & $0.5$ & $0.0212$ & $0.0701$ & $0.2521$ & $0.9769$ & $0.7587$
& $0.6983$ \\ 
& $3.5$ & $5.5$ & $1.5$ & $0.0940$ & $0.0615$ & $0.3568$ & $0.6310$ & $%
1.0365 $ & $0.9990$ \\ 
& $1$ & $1.5$ & $2$ & $0.0540$ & $0.0360$ & $0.2859$ & $0.3200$ & $1.1823$ & 
$0.4091$ \\ \hline
$500$ & $2.5$ & $1.5$ & $0.5$ & $0.0015$ & $0.0558$ & $0.0535$ & $0.1890$ & $%
0.1850$ & $0.3845$ \\ 
& $5$ & $2.5$ & $0.5$ & $0.0200$ & $0.0454$ & $0.1295$ & $0.6542$ & $0.4230$
& $0.4813$ \\ 
& $3.5$ & $5.5$ & $1.5$ & $0.0416$ & $0.0276$ & $0.1753$ & $0.4277$ & $%
0.8587 $ & $0.7053$ \\ 
& $1$ & $1.5$ & $2$ & $0.0329$ & $0.0177$ & $0.2083$ & $0.1981$ & $1.0690$ & 
$0.2688$ \\ \hline
$1000$ & $2.5$ & $1.5$ & $0.5$ & $0.0013$ & $0.0379$ & $0.0344$ & $0.1523$ & 
$0.1354$ & $0.3164$ \\ 
& $5$ & $2.5$ & $0.5$ & $0.0027$ & $0.0022$ & $0.0640$ & $0.5269$ & $0.3010$
& $0.3819$ \\ 
& $3.5$ & $5.5$ & $1.5$ & $0.0119$ & $0.0179$ & $0.1078$ & $0.3466$ & $%
0.6553 $ & $0.5455$ \\ 
& $1$ & $1.5$ & $2$ & $0.0152$ & $0.0130$ & $0.1341$ & $0.1585$ & $0.8015$ & 
$0.1926$ \\ \hline
\end{tabular}%
\end{table}%

\bigskip

\section{Real data analysis}

An application to well-known real data is presented to demonstrate the
flexibility of the PNGKME distribution. We took the data set from Lee and
Wang (2003), which consists of remission times, in months, of 128
individuals diagnosed with bladder cancer. The 128 data are: 0.08, 2.09,
3.48, 4.87, 6.94 , 8.66, 13.11, 23.63, 0.20, 2.23, 3.52, 4.98, 6.97, 9.02,
13.29, 0.40, 2.26, 3.57, 5.06, 7.09, 9.22, 13.80, 25.74, 0.50,2.46 , 3.64,
5.09, 7.26, 9.47, 14.24, 25.82, 0.51, 2.54, 3.70, 5.17, 7.28, 9.74,
14.76,26.31, 0.81, 2.62, 3.82, 5.32, 7.32, 10.06, 14.77, 32.15, 2.64, 3.88,
5.32, 7.39, 10.34,14.83, 34.26, 0.90, 2.69, 4.18, 5.34, 7.59, 10.66, 15.96,
36.66, 1.05, 2.69, 4.23, 5.41, 7.62, 10.75, 16.62, 43.01, 1.19, 2.75, 4.26,
5.41, 7.63, 17.12, 46.12, 1.26, 2.83, 4.33, 7.66, 11.25, 17.14, 79.05, 1.35,
2.87, 5.62, 7.87, 11.64, 17.36, 1.40, 3.02, 4.34, 5.71, 7.93, 11.79, 18.10,
1.46, 4.40, 5.85, 8.26, 11.98, 19.13, 1.76, 3.25, 4.50, 6.25, 8.37, 12.02,
2.02, 3.31, 4.51, 6.54, 8.53, 12.03, 20.28, 2.02, 3.36, 6.76, 12.07, 21.73,
2.07, 3.36, 6.93, 8.65, 12.63, 22.69, 5.49.\bigskip

The introduced PNGKME model was compared with its nested distributions (see
Section 2) and with the following distributions:\bigskip

\begin{itemize}
\item Exponentiated Weibull (EW) distribution (see Mudholkar and Srivastava,
1993) with PDF%
\begin{equation*}
f\left( x\right) =\alpha \beta \lambda x^{\lambda -1}e^{-\beta x^{\lambda
}}\left( 1-e^{-\beta x^{\lambda }}\right) ^{\alpha -1},\text{ }x>0,\text{ }%
\alpha >0,\text{ }\beta >0,\text{ }\lambda >0.
\end{equation*}

\item Weibull distribution with PDF%
\begin{equation*}
f\left( x\right) =\alpha \beta ^{-\alpha }x^{\alpha -1}e^{-\beta
^{-1}x^{\alpha }},\text{ \ }x>0,\text{ }\alpha >0,\text{ }\beta >0.
\end{equation*}

\item Gamma distribution with PDF%
\begin{equation*}
f\left( x\right) =\frac{\beta ^{\alpha }}{\Gamma \left( \alpha \right) }%
x^{\alpha -1}e^{-\beta x},\text{ \ }x>0,\text{ }\alpha >0,\text{ }\beta >0.
\end{equation*}
\end{itemize}

\noindent Table 2 lists the MLEs of all model parameters. The asymptotic
variance-covariance matrix of the MLEs is%
\begin{equation*}
\left( 
\begin{array}{ccc}
0.06141611 & 3.675929\times 10^{-5} & 0.001568761 \\ 
3.675929\times 10^{-5} & 2.200424\times 10^{-8} & 9.467161\times 10^{-7} \\ 
0.001568761 & 9.467161\times 10^{-7} & 6.190447\times 10^{-5}%
\end{array}%
\right) .
\end{equation*}

\noindent So, the approximate 95\% confidence intervals of the PNGKME model
parameters $\alpha ,\beta \ $and $\lambda \ $are

$\left( 1.14668\text{\ };\text{ }2.41793\right) ,$ $\left( 0.04908\text{\ };%
\text{\ }0.05615\right) $ and $\left( 41.18409\text{ };\text{ }%
41.18782\right) $ respectively.\medskip

To select the best model, we use the minus twice the maximized
log-likelihood ($-2\ln L$), Akaike information criterion (AIC),
Kolmogorov-Smirnov (K-S), Cramer-von-Mises (CvM), and Anderson-Darling (A-D)
test statistics with their $p$-values. We choose the best distribution,
which has the smallest values of these statistics and the largest $p$%
-values.\medskip

The statistics $-2\ln L$, and AIC are listed in Table 2. The values of the
K-S, CvM, and A-D statistics with their $p$-values are given in Table 3.
Therefore, we conclude that the proposed PNGKME distribution has a superior
fit compared to the other models.

\begin{table}[h]%
\caption{MLE, -2lnL and AIC  for fitted data. }$%
\begin{tabular}{lccccc}
\hline
\textbf{Distribution} & $\mathbf{\alpha }$ & $\mathbf{\beta }$ & $\mathbf{%
\lambda }$ & $\mathbf{-2}\ln \mathbf{L}$ & \textbf{AIC} \\ \hline
\textbf{PNGKME} & $1.6651$ & $0.05250$ & $41.1860$ & $\mathbf{820.1974}$ & $%
\mathbf{826.1974}$ \\ 
\textbf{GKME} & $-$ & $0.1113$ & $0.8539$ & $828.6188$ & $832.6188$ \\ 
\textbf{PKME} & $1.4330$ & $0.1034$ & $-$ & $823.0056$ & $827.0056$ \\ 
\textbf{KME} & $-$ & $0.0797$ & $-$ & $830.2841$ & $832.2841$ \\ 
\textbf{PAPTE} & $1.25611$ & $0.13042$ & $1.0069$ & $826.6821$ & $832.6821$
\\ 
\textbf{APTE} & $-$ & $1.1712$ & $0.1113$ & $828.6188$ & $832.6188$ \\ 
\textbf{PDUSE} & $1.0047$ & $0.1356$ & $-$ & $830.289$ & $834.289$ \\ 
\textbf{DUSE} & $-$ & $0.13527$ & $-$ & $830.2907$ & $832.2907$ \\ 
\textbf{PPETE} & $0.9763$ & $0.13747$ & $-$ & $830.9305$ & $834.9305$ \\ 
\textbf{PETE} & $-$ & $0.1394$ & $-$ & $830.9733$ & $832.9733$ \\ 
\textbf{EE} & $1.2174$ & $-$ & $-$ & $826.1486$ & $830.1486$ \\ 
\textbf{EW} & $2.7951$ & $0.2989$ & $0.6543$ & $821.3551$ & $827.3551$ \\ 
\textbf{Weibull} & $1.0476$ & $0.1046$ & $-$ & $828.1591$ & $832.1591$ \\ 
\textbf{Gamma} & $1.1717$ & $0.1251$ & $-$ & $826.7268$ & $830.7268$ \\ 
\textbf{Exponential} & $-$ & $0.1068$ & $-$ & $828.6646$ & $830.6646$ \\ 
\hline
\end{tabular}%
$%
\end{table}%

\bigskip

\begin{table}[h]%
\caption{K-S, CvM and A-D with their p-values. }$%
\begin{tabular}{lcccccc}
\hline
\textbf{Distribution} & \textbf{K-S} & \textbf{p-value} & \textbf{CvM} & 
\textbf{p-value} & \textbf{A-D} & \textbf{p-value} \\ \hline
\textbf{PNGKME} & $\mathbf{0.0386}$ & $\mathbf{0.9912}$ & $\mathbf{0.0278}$
& $\mathbf{0.9833}$ & $\mathbf{0.1952}$ & $\mathbf{0.9917}$ \\ 
\textbf{PKME} & $0.0564$ & $0.8099$ & $0.0707$ & $0.7482$ & $0.4215$ & $%
0.8269$ \\ 
\textbf{GKME} & $0.077751$ & $0.4214$ & $0.1656$ & $0.3457$ & $1.0813$ & $%
0.3172$ \\ 
\textbf{KME} & $0.1074$ & $0.1044$ & $0.3053$ & $0.1305$ & $1.9142$ & $%
0.1025 $ \\ 
\textbf{PAPTE} & $0.0525$ & $0.8716$ & $0.0613$ & $0.8068$ & $0.4858$ & $%
0.761$ \\ 
\textbf{APTE} & $0.077733$ & $0.4217$ & $0.16553$ & $0.3459$ & $1.0809$ & $%
0.3174$ \\ 
\textbf{PDUSE} & $0.0860$ & $0.3002$ & $0.1999$ & $0.2679$ & $1.0908$ & $%
0.3128$ \\ 
\textbf{DUSE} & $0.0856$ & $0.3048$ & $0.1978$ & $0.2719$ & $1.0853$ & $%
0.3153$ \\ 
\textbf{PPETE} & $0.0876$ & $0.2791$ & $0.2106$ & $0.2480$ & $1.1481$ & $%
0.2881$ \\ 
\textbf{PETE} & $0.0894$ & $0.2575$ & $0.2236$ & $0.2262$ & $1.1899$ & $%
0.2713$ \\ 
\textbf{EE} & $0.0725$ & $0.5124$ & $0.1273$ & $0.4673$ & $0.7105$ & $0.5498$
\\ 
\textbf{EW} & $0.0431$ & $0.9712$ & $0.0399$ & $0.9345$ & $0.2669$ & $0.9607$
\\ 
\textbf{Weibull} & $0.4023$ & $<2.2\times 10^{-16}$ & $9.0879$ & $<2.2\times
10^{-16}$ & $44.467$ & $4.687\times 10^{-6}$ \\ 
\textbf{Gamma} & $0.0732$ & $0.4986$ & $0.1349$ & $0.4394$ & $0.7683$ & $%
0.5042$ \\ 
\textbf{Exponential} & $0.0829$ & $0.3422$ & $0.1778$ & $0.3152$ & $1.1676$
& $0.2801$ \\ \hline
\end{tabular}%
$%
\end{table}%
\ 

\bigskip

\bigskip

\bigskip

\bigskip \newpage

\bigskip

\bigskip

\bigskip

\section{\textbf{Conclusions}}

Using the power generalization technique, we have proposed the PNGKM
distribution. This class contains ten transformations as special cases. We
have taken the exponential distribution as the parent model to obtain the
PNGKME distribution. The PNGKME distribution has eleven sub-models. Several
statistical properties of the PNGKME distribution are derived, like the rth
moment, generating, characteristic, and cumulant generating functions,
quantile function, mean deviations, mean residual life function, order
statistics, R\'{e}nyi entropy, and reliability. We have used the maximum
likelihood approach to estimate the unknown parameters of the PNGKME model.
In a simulation study, we have shown that the MLEs perform well. A famous
real data set is presented, and the results show that our new model provides
a better fit than the other distributions.\textbf{\medskip \medskip \medskip 
}

\noindent \textbf{Appendix.\medskip }

\noindent The observed information matrix of the parameters $\alpha $, $%
\beta $ and $\lambda $ has the following elements:%
\begin{align*}
\frac{\partial ^{2}\ln L}{\partial \alpha ^{2}}& =-\frac{n}{\alpha ^{2}},%
\text{ \ }\frac{\partial \ln L}{\partial \alpha \partial \beta }=-\ln
\lambda \sum_{i=1}^{n}\frac{x_{i}\lambda ^{e^{-\beta x_{i}}}e^{-\beta x_{i}}%
}{\lambda ^{e^{-\beta x_{i}}}-\lambda }, \\
\frac{\partial \ln L}{\partial \alpha \partial \lambda }& =\frac{n}{\lambda }%
-\frac{\alpha }{\lambda -1}+\sum_{i=1}^{n}\frac{\left( e^{-\beta
x_{i}}-1\right) \lambda ^{-\left( 2-e^{-\beta x_{i}}\right) }}{1-\lambda
^{-\left( 1-e^{-\beta x_{i}}\right) }}, \\
\frac{\partial ^{2}\ln L}{\partial \beta ^{2}}& =-\frac{n}{\beta ^{2}}%
+\sum_{i=1}^{n}\frac{x_{i}^{2}\lambda ^{2e^{-\beta x_{i}}}e^{-2\beta
x_{i}}\left( 1-\alpha \right) \left( \ln \lambda \right) ^{2}}{\left(
\lambda ^{e^{-\beta x_{i}}}-\lambda \right) ^{2}}+\left( \ln \lambda \right)
\sum_{i=1}^{n}x_{i}^{2}e^{-\beta x_{i}} \\
& -\sum_{i=1}^{n}\frac{x_{i}^{2}\lambda ^{e^{-\beta x_{i}}}e^{-2\beta
x_{i}}\left( 1-\alpha \right) \left( \ln \lambda \right) ^{2}}{\lambda
^{e^{-\beta x_{i}}}-\lambda }-\sum_{i=1}^{n}\frac{x_{i}^{2}\lambda
^{e^{-\beta x_{i}}}e^{-\beta x_{i}}\left( 1-\alpha \right) \ln \lambda }{%
\lambda ^{e^{-\beta x_{i}}}-\lambda }, \\
\frac{\partial \ln L}{\partial \beta \partial \lambda }& =\sum_{i=1}^{n}%
\frac{x_{i}^{2}e^{-2\beta x_{i}}\left[ \left( \left( 1-\alpha \right)
e^{-\beta x_{i}}-\alpha +1\right) \lambda ^{e^{-\beta x_{i}}+1}\ln \left(
\lambda \right) \right] }{\lambda \left( \lambda ^{e^{-\beta x_{i}}}-\lambda
\right) ^{2}} \\
& -\sum_{i=1}^{n}\frac{\left( \alpha +1\right) e^{-\beta x_{i}}\lambda
^{e^{-\beta x_{i}}+1}+\alpha e^{-\beta x_{i}}\lambda ^{2e^{-\beta
x_{i}}}+\lambda ^{2}e^{-\beta x_{i}}}{\lambda \left( \lambda ^{e^{-\beta
x_{i}}}-\lambda \right) ^{2}},
\end{align*}%
and%
\begin{align*}
\frac{\partial ^{2}\ln L}{\partial \lambda ^{2}}& =-\frac{\alpha n}{\left(
\lambda -1\right) ^{2}}+\frac{n\left( 1+\ln \lambda \right) }{\lambda
^{2}\left( \ln \lambda \right) ^{2}}+\sum_{i=1}^{n}\frac{\left( \alpha
-1\right) \left( e^{-\beta x_{i}}-1\right) ^{2}\lambda ^{-4+2e^{-\beta
x_{i}}}}{\left( 1-\lambda ^{-\left( 1-e^{-\beta x_{i}}\right) }\right) ^{2}}
\\
& -\sum_{i=1}^{n}\frac{e^{-\beta x_{i}}+n\left( \alpha -1\right) }{\lambda
^{2}}+\sum_{i=1}^{n}\frac{\left( \alpha -1\right) \left( e^{-\beta
x_{i}}-2\right) \left( e^{-\beta x_{i}}-1\right) \lambda ^{-3+e^{-\beta
x_{i}}}}{1-\lambda ^{-\left( 1-e^{-\beta x_{i}}\right) }}.
\end{align*}

\newpage

\bigskip \bigskip \bigskip

\noindent \textbf{Funding} There was no funding for the research.\bigskip
\bigskip \bigskip

\noindent \textbf{Declarations }\smallskip \bigskip

\noindent \textbf{Conflict of interest }There is no Conflict of
interest.


\bigskip \bigskip \bigskip \bigskip

\begin{thebibliography}{99}
\bibitem{} Arnold, BC., Balakrishnan N, Nagarajah HN. (2008) A first course
in order statistics. New York: John Wiley.

\bibitem{} Burnham KP (1987) Design and analysis methods for fish survival
experiments based on release-recapture. American Fisheries Society,
Monograph 5:1--437

\bibitem{} Deepthi K. S, Chacko V. M. (2023). Power Generalized
KM-Transformation for Non-Monotone Failure Rate Distribution.
https://arxiv.org/abs/2308.14770.

\bibitem{} Gupta RC, Gupta PL, Gupta RD (1998) Modeling failure time data by
lehman alternatives. Communications in Statistics-Theory and methods
27(4):887--904.

\bibitem{} Kavya P, Manoharan M (2021) Some parsimonious models for
lifetimes and applications. J Stat Comput Simul 91(18):3693--3708

\bibitem{} Kumar D, Singh U, Singh SK (2015) A method of proposing new
distribution and its application to bladder cancer patients data. J. Stat.
Appl. Pro. Lett 2(3):235--245.

\bibitem{} Lee, E. T. and Wang, J. W. (2003): Statistical Methods for
Survival Data Analysis. Wiley, New York, DOI:10.1002/0471458546.

\bibitem{} Lone, M. A., Jan, T. R., 2023, A new Pi-exponentiated method for
constructing distributions with an application to Weibull distribution,
Reliability: Theory \& Applications, 18(1 (72)):94--109,

\bibitem{} Mahdavi A, Kundu D (2017) A new method for generating
distributions with an application to exponential distribution.
Communications in Statistics-Theory and Methods 46(13):6543--6557

\bibitem{} Mudholkar GS, Srivastava DK (1993) Exponentiated weibull family
for analyzing bathtub failure-rate data. IEEE Trans Reliab 42(2):299--302

\bibitem{} Nadarajah S (2011) The exponentiated exponential distribution: a
survey. AStA Advances in Statistical Analysis 95:219--251

\bibitem{} Nadarajah S, Kotz S (2006) The beta exponential distribution.
Reliability engineering \& system safety 91(6):689--697.

\bibitem{} R\'{e}nyi, A. (1961). On measures of entropy and information.
Proceedings of the fourth berkeley symposium on mathematical statistics and
probability, volume 1: Contributions to the theory of statistics (Vol. 4,
pp. 547-562)

\bibitem{} Shannon CE (1951) Prediction and entropy of printed english. Bell
Syst Tech J 30(1):50--64

\bibitem{} Sharma VK, Singh SK, Singh U, Merovci F (2016) The generalized
inverse lindley distribution: A new inverse statistical model for the study
of upside- down bathtub data. Communications in Statistics-Theory and
Methods 45(19):5709--5729

\bibitem{} Singh SK, Singh U, Sharma VK (2013) Expected total test time and
bayesian estimation for generalized lindley distribution under progressively
type-ii cen- sored sample where removals follow the beta-binomial
probability law. Appl Math Comput 222:402--419

\bibitem{} Thomas, B., Chacko, V. M. (2021). Power Generalized DUS
Transformation of Exponential Distribution. International Journal of
Statistics and Reliability Engineering, 9(1), 150-157.

\bibitem{} Verma E, Kumar S.S., Yadav S. (2025) A new generalized class of
Kavya--Manoharan distributions: inferences and applications. Life Cycle
Reliability and Safety Engineering 14:79--91.
\end{thebibliography}
\end{document}